\documentclass{sigchi}
\pdfoutput=1
\usepackage{graphicx}
\usepackage[]{color}

\makeatletter
\def\maxwidth{ %
  \ifdim\Gin@nat@width>\linewidth
    \linewidth
  \else
    \Gin@nat@width
  \fi
}
\makeatother

\definecolor{fgcolor}{rgb}{0.345, 0.345, 0.345}

\usepackage{framed}
\makeatletter
 {\par\unskip\endMakeFramed%
 \at@end@of@kframe}
\makeatother

\definecolor{shadecolor}{rgb}{.97, .97, .97}
\definecolor{messagecolor}{rgb}{0, 0, 0}
\definecolor{warningcolor}{rgb}{1, 0, 1}
\definecolor{errorcolor}{rgb}{1, 0, 0}

\usepackage{alltt}

\usepackage[
  pass,
]{geometry}

\usepackage{balance}
\usepackage{amsmath}
\usepackage{graphicx}
\usepackage{booktabs}
\usepackage{colortbl}
\usepackage[usenames,dvipsnames]{xcolor}
\usepackage{bm}
\usepackage{url}
\usepackage{booktabs}
\usepackage{textcomp}
\usepackage{microtype} 
\usepackage{ccicons}

\usepackage{ucs}
\usepackage[utf8x]{inputenc}

\usepackage[]{color}
\usepackage[breaklinks]{hyperref}

\hypersetup{colorlinks=true, linkcolor=Black, citecolor=Black, filecolor=Blue,
    urlcolor=Blue, unicode=true}

\usepackage[all]{hypcap}  
\IfFileExists{upquote.sty}{\usepackage{upquote}}{}
\begin{document}

\conferenceinfo{CSCW}{'16 San Francisco, California, USA}

\title{Remixing as a Pathway to Computational Thinking}

\numberofauthors{4}
\author{
  \alignauthor Sayamindu Dasgupta\\
    \affaddr{MIT Media Lab}\\
    \affaddr{Cambridge, MA 02142}\\
    \email{sayamindu@media.mit.edu}\\
  \alignauthor William Hale\\
    \affaddr{University of Washington}\\
    \affaddr{Seattle, WA 98195}\\
    \email{halew@uw.edu}\\
  \alignauthor Andres Monroy-Hernandez\\
    \affaddr{Microsoft Research}\\
    \affaddr{Redmond, WA 98052}\\
    \email{amh@microsoft.com}\\
  \alignauthor Benjamin Mako Hill\\
    \affaddr{University of Washington}\\
    \affaddr{Seattle, WA 98195}\\
    \email{makohill@uw.edu}
}

\maketitle
\begin{abstract}
Theorists and advocates of ``remixing'' have suggested that appropriation can act as a pathway for learning. We test this theory quantitatively using data from more than 2.4 million multimedia programming projects shared by more than 1 million users in the Scratch online community. First, we show that users who remix more often have larger repertoires of programming commands even after controlling for the numbers of projects and amount of code shared. Second, we show that exposure to computational thinking concepts through remixing is associated with increased likelihood of using those concepts. Our results support theories that young people learn through remixing, and have important implications for designers of social computing systems.
\end{abstract}




\section{Introduction}
When the Lifelong Kindergarten group at MIT designed and built an online community for users of the Scratch programming language in 2006 and 2007, the system was designed as a platform to help young people learn to program through remixing \cite{monroy-hernandez_scratchr:_2007}. Every project shared publicly on the Scratch website is released under a license -- explained in child-friendly terms -- that allows unrestricted reuse and modification. Not an easy or obvious decision at the time, the MIT team embraced remixing with the hope that novice users might achieve their goals by downloading, studying, and modifying existing projects created by more skilled users, and through that process, they would learn programming skills.

\begin{figure}[t!]
  \centering
      \includegraphics[width=0.8\columnwidth]{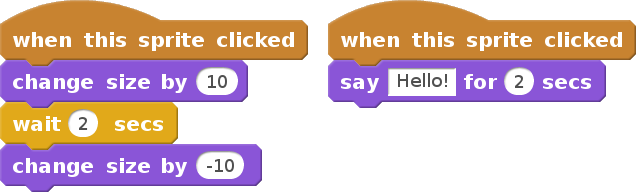}
  \caption{Scratch code that will cause clicking on a visual object (i.e., sprite) to grow in size, and then shrink, while showing a speech bubble at the same time.
  }
\label{fig:scratchcode}
\end{figure}

The idea that remixing and appropriation of content can promote learning has broad currency in the literatures on social computing and youth and media. Appropriation or remixing has been described as an important literacy or skill for young people by Bruckman \cite{bruckman_community_1998}, Jenkins \cite{jenkins_confronting_2006}, Ito \cite{ito_hanging_2009}, Lessig \cite{lessig_remix:_2008}, Manovich \cite{manovich_remix_2005}, and others. One of the most important arguments made in favor of remixing is that it allows users to engage with material created by others with different skills, knowledge, and experiences, and that this exposure promotes learning. The idea that remixing can promote learning has been particularly influential in studies of programming. Scratch designers' embrace of ``creative appropriation'' \cite{monroy-hernandez_scratchr:_2007} is only one example of a strong current in constructionist approaches to learning \cite{papert_mindstorms:_1980} that attempt to place the learning of ``computational thinking'' concepts \cite{wing_computational_2006,wing_computational_2008} in a social context \cite{papert_poetic_1976, bruckman_community_1998}.

This paper empirically tests the theory that young programmers can increase their skills and learn computational thinking concepts through remixing other programmers' code. Using a longitudinal dataset of 2,426,894 projects created by 1,068,502 users over the first five years of the Scratch online community, we examine the association between engagement in remixing and the level and speed with which users demonstrate new computational thinking concepts. Although limited in several ways, our results provide broad support for the idea that young people learn through remixing.

\section{Background}

Remixing has been defined as the reworking and combination of existing creative artifacts, usually in the form of music, video, and other interactive media. The phenomenon is widespread, culturally significant, and controversial. Lessig has suggested that remixing reflects a broad cultural shift spurred by the Internet and a source of enormous creative potential \cite{lessig_remix:_2008}. Manovich has called remixing ``a built-in feature of the digital networked media universe'' \cite{manovich_remix_2005}. Benkler has cited remixing as an example of the power and potential of a new mode of social production \cite{benkler_wealth_2006}.

Remixing online is described as important for at least three reasons. First, theorists like Lessig \cite{lessig_remix:_2008} take a normative position that participation in cultural production is \emph{prima facie} socially beneficial and have suggested that remixing represents a low-cost and accessible form of participation. Second, scholars like Benkler \cite{benkler_wealth_2006} have pointed to remixing as a possible path toward high quality information goods through the mass aggregation of many small contributions in ways that are similar, in both process and effect, to Wikipedia or free/open source software. Finally, theorists including Jenkins \cite{jenkins_confronting_2006} have advanced an argument that remixing leads to learning as a form of legitimate peripheral participation \cite{lave_situated_1991}. Of course, remixing has not been without its detractors. Keen \cite{keen_cult_2007} and Lanier \cite{lanier_you_2010} have both suggested that remixing systematically falls short of its promise, in all three senses, and that remixes are largely derivative, uninteresting, and of poor quality.

Driven by excitement about the promise of remixing, a large body of empirical research has been generated on the subject. Much of this work has focused on mapping dynamics within remixing communities to understand why some users engage in remixing \cite{cheliotis_analysis_2009, yew_social_2009} or why some artifacts are remixed while most never become sites for collaboration \cite{hill_remixing_2013, cheliotis_antecedents_2014, oehlberg_patterns_2015}. Because remixing involves appropriation, another body of work has looked at authorship and credit-giving \cite{luther_edits_2010, monroy-hernandez_computers_2011, kim_appropriate_2015} and to intersections between remixing practice and copyright law \cite{lessig_remix:_2008, hemphill_remix_2009, seneviratne_remix_2010, fiesler_remixers_2014}. In line with theory, empirical work has also considered remixing and its relationship to quality in order to explore if, and when, remixing leads to better results than what is achieved by creators working alone \cite{yu_cooks_2011, hill_cost_2013}.

While empirical remixing research has explored means of supporting and promoting remixing, tested the reception of remixes by users, and tested theories about the quality of remixed outputs, we know of no research that has empirically tested the theory that remixing acts a pathway to learning. That said, the idea that remixing can promote learning has been influential to designers of remixing systems. The creators of the Scratch online community cited Jenkins \cite{jenkins_convergence_2008} to describe their ideal model of ``active engagement'' with content, where ``members of the community can share or appropriate the original building blocks of the Scratch projects they interact with,'' and where users of the Scratch programming language can learn through remixing others' content \cite{monroy-hernandez_scratchr:_2007}. Outside of remixing communities, research on professional software engineers has documented examples of appropriation as a pathway to learning \cite{brandt_two_2009, hartmann_hacking_2008}.

The idea of a community that would use remixing to support learning about programming finds particular support in the theory of constructionism \cite{papert_mindstorms:_1980, resnick_pianos_1996} that motivated the design of Scratch itself \cite{resnick_scratch:_2009}. Constructionist theories hold that we learn best by designing and constructing ``public entities,'' and that this learning is even more meaningful and effective in a social context \cite{papert_poetic_1976}. As a result, research on learning in Scratch has focused on social interactions and relationships within the community \cite{brennan_more_2011,brennan_imagining_2013,fernando_online_2014} as well as on appropriation within the Scratch website \cite{hill_responses_2010,monroy-hernandez_computers_2011,hill_cost_2013,hill_remixing_2013}.

Although users of Scratch learn a wide range of 21st century literacies \cite{jenkins_confronting_2006}, the system was designed to support young people in learning to program. While constructionism has been primarily focused on epistemology and ways of learning, other research on computational thinking (CT) \cite{wing_computational_2006,wing_computational_2008} has influenced and described many of the concepts that Scratch hopes its users will learn. Wing, who coined the term, defines computational thinking as ``the thought processes involved in formulating problems and their solutions so that the solutions are represented in a form that can be effectively carried out by an information-processing agent'' \cite{wing_computational_2010}.
Recent work by Brennan and Resnick \cite{brennan_new_2012} has unpacked CT in terms of concepts, practices, and perspectives. In this treatment, concepts reflect the most basic building blocks of computation (like sequences, loops, and conditionals) while practices and perspectives involve higher level strategies and world-views (like debugging, or reflections on the learning process itself).

\begin{figure*}[t!]
\centering
\includegraphics[width=1.0\textwidth]{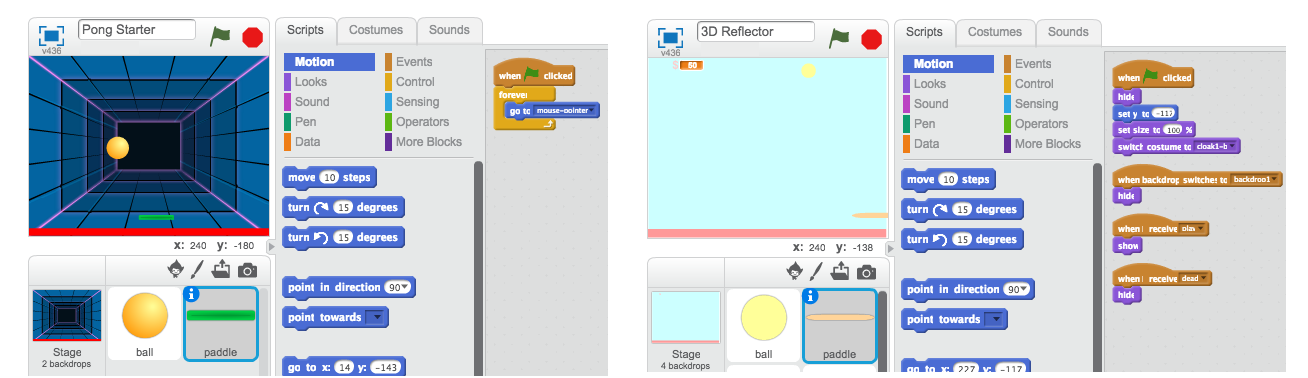}
\caption{Remix of a Scratch ``pong'' game (original project on the left), with modification of graphic elements, as well as more sophisticated code for added functionality.}
\label{fig:scratch_remix}
\end{figure*}

Our goal is to test the theory that remixing or appropriation of code is a mechanism through which individuals can be exposed to and learn computational thinking concepts. Attempts to measure learning are always challenging and controversial and doing so in informal learning environments is even more difficult. Several previous studies have looked at learning in Scratch, although most have been case-study based and qualitative \cite{maloney_programming_2008, dahotre_qualitative_2010, peppler_uncovering_2011, scaffidi_skill_2011, brennan_best_2013}. We are particularly inspired by recent work by Yang et al. \cite{yang_uncovering_2015} that improves on previous quantitative studies of learning in informal environments by looking at user ``trajectories.''

Yang et al.\ models learning as growth in the cumulative repertoire of programming tokens, similar to a measure of demonstrated vocabulary, which may grow more or less quickly over time. Building on this approach, we seek to examine how, \emph{ceteris paribus}, a learner's repertoire of programming concepts increases when she engages in remixing others' projects that use unfamiliar blocks. This leads us to our first hypothesis (H1): \emph{changes in a user's programming repertoire will be larger when she has engaged in more remixing activity.}

We are also inspired by work by Scaffidi and Chambers who seek to measure learners' ``breadth'' in terms of the number of types of concepts a user has demonstrated \cite{scaffidi_skill_2011}, and by Brennan and Resnick's description of particular CT concepts \cite{brennan_new_2012}. As a result, we offer a narrower group of hypotheses focused on the relationship between remixing and individual CT concepts. Our second hypothesis (H2) is that \emph{a user who remixes more projects that use a particular computational thinking concept will be more likely to use that concept for the first time in a de novo project than a user who has remixed fewer of these projects.} We test this theory as it applies to each of Brennan and Resnick's CT concepts separately.

\section{Empirical Setting}

The current work contributes to the growing body of previous work in social computing on remixing and learning using the Scratch online community as its empirical setting. Scratch is a public and freely accessible website\footnote{\url{https://scratch.mit.edu}} where a large community of users create, share, and remix interactive media with the Scratch programming environment.
With more than 7 million users, 10 million projects, and 2.6 million remixes, 
Scratch is the largest community dedicated to remixing as well as the largest informal community for young people to learn programming.

The Scratch programming environment is visual, block-based, and designed for novice users \cite{resnick_scratch:_2009}. Programs in Scratch are constructed by dragging and dropping visual blocks together (see Figures \ref{fig:scratchcode} and \ref{fig:scratch_remix}). Blocks are analogous to tokens or symbols in the source code of traditional computer programs, and can be used to do things like update a variable, move an object on the screen, play a sound, or repeat a sequence of other blocks. An analogy can be drawn between blocks and lines of code, though blocks are more granular. Scratch includes more than 120 distinct blocks which can be combined together to form ``scripts.'' Figure \ref{fig:scratchcode}, for example, shows two scripts with 4 and 2 blocks side-by-side.

Blocks define the behavior of on-screen graphical objects called ``sprites''  which can interact with each other and the user. Projects on the Scratch website vary enormously in subject matter as well as in complexity in terms of both code and media. Visitors to the Scratch website must create accounts to share projects or to contribute in other ways such as commenting, showing support (i.e., giving ``loveits''), tagging, or flagging projects as inappropriate. Most of the community’s users self-report their ages ranging between 8 and 17 with 12 being the median age for new accounts. As of May 2013 the Scratch programming environment has been fully integrated into the web browser and online community. Our analysis covers the Scratch community during a window from marzo -d, 2007 to abril -d, 2012, before the introduction of the web-based editing environment.

\section{Data and Measures}
\begin{table*}[ht]
  \centering
  \begin{tabular}{lp{1.4in}p{4.5in}}
    \toprule
    Concepts & Measure & Scratch Blocks \\
    \midrule
    Loops & Uses looping blocks (e.g., Forever block) & forever, foreverIf, repeat, repeatUntil\\
    \\
    Parallelism & Parallel scripts with same ``hat'' block. & startHatTriggered, eventHatTriggered, keyHatTriggered, mouseHatTriggered\\
    \\
    Events & Uses ``when $<>$'' hat blocks & eventHatTriggered, keyHatTriggered, mouseHatTriggered, bounceOffEdge, turnAwayFromEdge, touching, touchingColor, colorSees, mousePressed, keyPressed, isLoud, sensor, sensorPressed, distanceTo\\
    \\
    Conditionals & Uses conditional blocks (e.g. ``if'' block) & waitUntil, foreverIf, if, ifElse, repeatUntil, bounceOffEdge, turnAwayFromEdge, touching, touchingColor, colorSees, mousePressed, keyPressed, isLoud, sensor, sensorPressed, lessThan, equalTo, greaterThan, and, or, not, listContains\\
    \\
    Operators & Uses operator blocks (e.g. ``+'' or ``or'' blocks) & lessThan, equalTo, greaterThan, and, or, not, add, subtract, multiply, divide, pickRandomFromTo, concatenateWith, letterOf, stringLength, mod, round, abs, sqrt, sin, cos, tan, asin, acos, atan, ln, log, e\textasciicircum, 10\textasciicircum\\
    \\
    Data & Uses data blocks (e.g. Variable block) & setVarTo, changeVarBy, showVariable, hideVariable, readVariable, addToList, deleteLineOfList, insertAtOfList, setLineOfListTo, contentsOfList, getLineOfList, lineCountOfList, listContains\\
    \bottomrule
  \end{tabular}
  \caption{Mapping of CT concepts to Scratch blocks.}
  \label{tab:concepts}
\end{table*}

The Scratch online community uses a database-driven web application that stores an extensive range of metadata on projects, users, and interactions on the website. This database also identifies, tracks, and presents data on whether projects are created through remixing. Additionally, the website stores each of the raw Scratch project files which can be further analyzed to reveal details such as projects' programming code and media elements. Our dataset is constructed by combining exported metadata about Scratch’s users, projects, and interactions with algorithmic analyses of each project.
Our unit of analysis is the Scratch project and our dataset includes every project shared on the Scratch online community from the moment the first project was shared in Scratch on marzo -d, 2007 through abril -d, 2012 -- a total of 2,426,894 projects.

Because our study aims to measure learning by looking at within-user changes, our analysis focuses on the 173,053 users who have shared at least two projects. We mark age data as missing for 4,354 users who report their age (at the time that they shared their first project) as less than 4 or more than 90 years. Finally, we omit projects for which we do not have data because of technical errors in our analytic tools or because of corruption in the project files. For analyzing overall repertoire size, we consider 1,625,988 de novo projects shared by all users within our restricted dataset, out of which we omit 4,059 projects due to missing data. While analyzing for specific concepts, we omit 4,950 projects due to missing data and analyze a total of 2,280,709 projects shared or remixed by the same set of users. For parallelism, we omit 11,229 projects due to technical errors in our analytic tools or missing or corrupt data and analyze a total of 2,274,435 projects.

To test hypothesis H1 about growth in programming repertoire, we operationalize a user's repertoire of programming concepts as the number of programming block types that they have ever used. To test hypothesis H2, we measure the presence or absence of particular computational thinking concepts. Toward this end, we adopt Brennan and Resnick's mapping of Scratch blocks to CT concepts \cite{brennan_new_2012} as described in Table \ref{tab:concepts}. For example, use of the ``repeat'' block in Scratch is one indicator of the use of the CT concept of \emph{loops}. To measure the CT concept of ``parallelism,'' we parse projects' code structure to detect multiple event-handling blocks that are listening for the same event. Figure \ref{fig:scratchcode} is a minimal example of Scratch code expressing parallelism with two instances of the same event-handler block with different blocks attached to each.
We deviate from Brennan and Resnick only in that we do not attempt to measure the CT concept of ``sequences'' as any block connected to another block is a sequence and nearly every Scratch project includes this.

\begin{table}[t!]
\centering
\begin{tabular}{rrrrl}
  \toprule
 & {M} & {$\bar{x}$} & {$\sigma$} & {Range} \\ 
  \midrule
Cum. Repertoire & 23 & 28 & 21 & [0,142] \\ 
  Remixes & 0 & 3 & 15 & [0,1008] \\ 
  De Novo Projects & 4 & 13 & 46 & [1,6347] \\ 
  Comments & 0 & 53 & 601 & [0,64374] \\ 
  User Age (Years) & 14 & 18 & 10 & [4,90] \\ 
  Experience (Days) & 13 & 99 & 206 & [0,1829] \\ 
  Total Blocks & 146 & 767 & 3402 & [0,525964] \\ 
  Downloads & 1 & 19 & 99 & [0,5591] \\ 
   \bottomrule
\end{tabular}
\caption{Summary statistics for measures of users at the point of data collection. Columns are included for the median (M). mean ($\bar{x}$), standard deviation ($\sigma$) and range.} 
\label{tab:sum}
\end{table}

Because our behavioral measures of learning may be influenced by a wide range of factors other than the amount of remixing a user has engaged in, we also include a range of control variables. Most critically, we include a count of the number of de novo projects a user has shared (\emph{De Novo Projects}) which provides the independent variable for modeling growth in repertoire associated with increased experience.

For some users of Scratch, interaction and sharing projects is primarily a social activity. Because these users' repertoires may grow more slowly, we include a control for the number of comments received by a user (\emph{Comments}); we expect a negative relationship between this measure and block repertoire. Because learning is related to development in general, we also include a self-reported measure of age in years at the moment that each project was shared (\emph{User Age}) and the age of each account in days (\emph{Experience}). Both of these variables may capture the sophistication of the user and we expect a positive relationship between these two variables and block repertoire.

In designing our study, we were concerned by several active sub-communities within Scratch that are characterized by large amounts of remixing and small amounts of project code. For example, many Scratch users engage in Scratch primarily through ``coloring-contests'' where users use remixing to modify image media in a given project \cite{nickerson_appropriation_2011}. Since these projects tend to have few programming blocks, we were concerned that this type of remixing might bias our estimate of the relationship between remixing through exposure to code. To account for this, we constructed a control for the aggregate corpus size across each user's de novo projects (\emph{Total Blocks}).

\begin{figure}[t]
\includegraphics[width=\columnwidth]{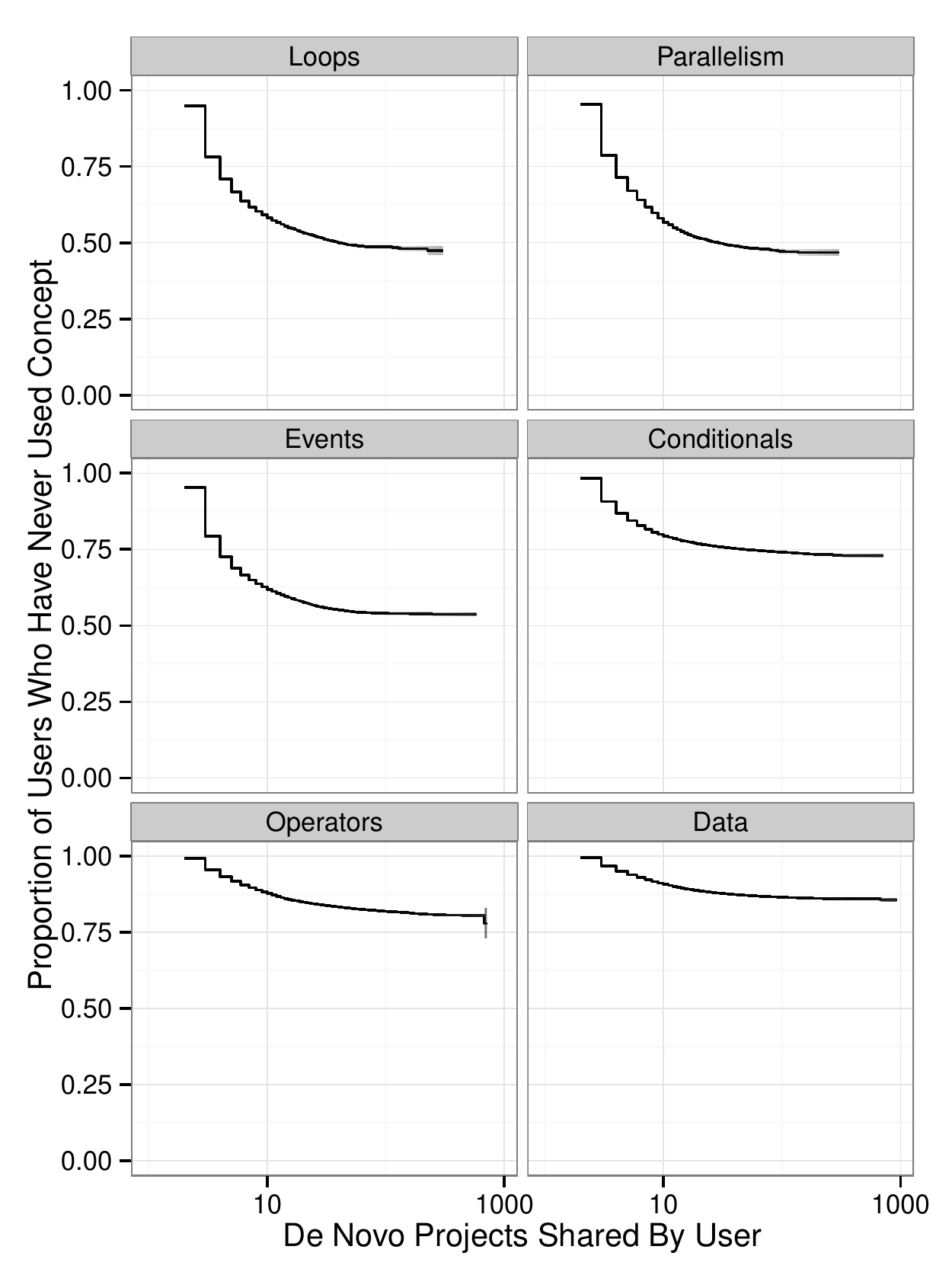}
\caption{Kaplan-Meier ``survival'' curves for users in Scratch. The $y$ axes show the portion of users who have never uploaded a \emph{de novo} project using a given programming concept. The $x$ axes show ``time'' in terms of the number of \emph{de novo} projects. }
\label{fig:km}
\end{figure}

For our final control measure, we use the number of projects a user has downloaded (\emph{Downloads}). \emph{Downloads} is a complicated measure for several reasons. During the timespan our dataset covers, the only way to view the source code of a project shared on Scratch was to download it. If a user downloads a project, learns how to use a new block, but does not \emph{share} a modified version of the downloaded project publicly, it will not be counted in \emph{Remixes} in our dataset. That said, exposure to and reuse of a block through downloading would constitute ``remixing'' as defined by Manovich \cite{manovich_remix_2005} and as appropriation as defined by Papert \cite{papert_mindstorms:_1980}. In this sense, the relationship between downloads and measures of learning may itself reflect evidence of learning through remixing. Although our analysis focuses on the more conservative measure of remixing used within the Scratch community, we urge readers to interpret our results in this context.

Table \ref{tab:sum} includes summary statistics for each of our variables at the user level at the end of our period of data collection on abril -d, 2012. As is typical for data on participation in online communities, nearly all the variables are highly skewed. To assist in interpretation and the satisfaction of parametric assumptions in our models, we apply a log-transformation to each of our independent variables to ensure that their distribution of values more closely approximates normality.

\section{Analysis}

For H1, we analyze longitudinal data that represents the block repertoire of each Scratch user, use projects as our unit of analysis, and include an observation for every de novo project shared in Scratch during our window of data collection by a user with two or more projects. To model the effects of remixing within users, we use panel regression models with user level fixed effects that are equivalent to fitting a dummy variable for every user in our sample \cite{singer_applied_2003}. Our formal model to test H1 is shown below:
\begin{equation*}
\begin{split}
\mathrm{Cumulative~Repertoire} = \beta_{0} + \beta_{1}\log{\mathrm{Remixes}} \\
                        + \beta_{2}\log{\mathrm{De~Novo~Projects}} + \beta_{3}\log{\mathrm{Comments}} \\
                        + \beta_{4}\log{\mathrm{User~Age}} + \beta_{5}\log\mathrm{Experience} \\
                        + \beta_{6}\log{\mathrm{Total~Blocks}} + \beta_{7}\log{\mathrm{Downloads}} \\
                        + \boldsymbol{\beta_{u}}\mathbf{USER}\boldsymbol{_u} + \varepsilon
\end{split}
\end{equation*}
To test our theory about the effect of coloring-contests, we fit an exploratory model (M1) without controlling for \emph{Total Blocks} or \emph{Downloads}, a model (M2) with the \emph{Total Blocks} control, and a model (M3) which also adds the \emph{Downloads} control.

To test H2, we model the effect of remixing projects that contain particular CT concepts on the likelihood of the user employing those concepts in subsequent de novo projects. The demonstration of concepts can happen at any time during the tenure of a Scratch user and is, as a result, well suited to modeling using a continuous time-survival approach using the Cox proportional hazards model \cite{singer_applied_2003}. Originally created for epidemiology models of actual human survival against disease, the language of these models can seem strange when applied to positive events like learning. To support a Cox proportional hazards approach, we modified our dataset so that each observation represents a spell between projects.

Figure \ref{fig:km} includes non-parametric plots that show the proportion of users who have never used a particular block as a function of the number of projects they have shared. Users contribute data until they either use the concept, or they are censored because they did not contribute additional projects. Cox models estimate the association between predictor variables and the ``risk'' of experiencing an event as multipliers of a baseline ``hazard'' function that is itself a function of the ``survival'' functions shown in Figure \ref{fig:km}.

Rather than treating time as calendar time, which produced substantively similar estimates, we use the number of projects shared as our baseline time variable $t$. The independent variable in our Cox models is the number of remixes a user has shared that use the CT concept in question (\emph{Remixes w/ Concept}). We also include \emph{Remixes}, capturing total remixing activity as a control alongside the other control variables used in our repertoire models. We model the risk of a user demonstrating a new CT concept, having shared a given number of projects $\lambda(t)$, as a function of a baseline hazard function $\lambda_{0}(t)$ in the following model:
\begin{equation*}
\begin{split}
\lambda(t\vert X) = \lambda_{0}(t) \exp\{ \beta_{1}\log{\mathrm{Remixes~w/~Concept}} \\
    + \beta_{2}\log{\mathrm{Remixes}} + \beta_{3}\log{\mathrm{De~Novo~Projects}} \\
    + \beta_{4}\log{\mathrm{Comments}} + \beta_{5}\log{\mathrm{User~Age}} \\
    + \beta_{6}\log\mathrm{Experience} + \beta_{7}\log{\mathrm{Total~Blocks}} \\
    + \beta_{8}\log{\mathrm{Downloads}} \}
\end{split}
\end{equation*}

We estimate a separate model for each CT concept: conditionals, data, events, loops, operators, and parallelism.

\section{Results}

\begin{table}
\begin{center}

\begin{tabular}{l c c c }
\hline
                    & M1 & M2 & M3 \\
\hline
ln Remixes          & $-0.304^{*}$ & $1.402^{*}$ & $0.253^{*}$  \\
                    & $(0.017)$    & $(0.016)$   & $(0.017)$    \\
ln De Novo Projects & $13.073^{*}$ & $4.779^{*}$ & $4.750^{*}$  \\
                    & $(0.023)$    & $(0.026)$   & $(0.026)$    \\
ln Comments         & $0.413^{*}$  & $0.643^{*}$ & $0.080^{*}$  \\
                    & $(0.011)$    & $(0.010)$   & $(0.011)$    \\
User Age            & $2.367^{*}$  & $2.912^{*}$ & $2.517^{*}$  \\
                    & $(0.019)$    & $(0.018)$   & $(0.018)$    \\
ln Experience       & $0.862^{*}$  & $0.128^{*}$ & $-0.046^{*}$ \\
                    & $(0.008)$    & $(0.007)$   & $(0.007)$    \\
ln Total Blocks     &              & $5.965^{*}$ & $5.823^{*}$  \\
                    &              & $(0.011)$   & $(0.011)$    \\
ln Downloads        &              &             & $2.237^{*}$  \\
                    &              &             & $(0.012)$    \\
\hline
R$^2$               & 0.751        & 0.793       & 0.798        \\
Adj. R$^2$          & 0.673        & 0.711       & 0.715        \\
Num. obs.           & 1578362      & 1578362     & 1578362      \\
\hline
\multicolumn{4}{l}{\scriptsize{$^*p<0.001$}}
\end{tabular}

\caption{Fitted regression models for Scratch users' cumulative block repertoires. All models use user-level fixed effects and reflect within-user estimates.}
\label{tab:rep}
\end{center}
\end{table}

Our tests of H1 and the relationship between remixing and blocks in a user's repertoire are shown in Table \ref{tab:rep}. In our first exploratory model M1, we find a negative association between the number of remixes a user has shared and the size of their repertoire. Controlling only for the total number of projects, comments, and users' age and experience, a one log-unit increase in the number of remixes a user shares is associated with a within-user decrease of 0.3 distinct blocks in her repertoire. Of course, this negative effect may simply reflect the fact that many users engaged in remixing are engaged in coloring-contests that involve little or no code. Model M2 attempts to address this concern by adding our control for the total number of blocks shared. Once we control for total block use, we find the parameter estimate for \emph{Remixes} is switched in sign and much larger. In this expanded model, we estimate that a one log-unit increase in number of remixes shared is associated with a 1.4 block increase in repertoire.

M3 adds a control variable for \emph{Downloads} and provides confirming evidence for our prediction that downloading -- a step toward remixing in Scratch -- moderates the effect of remixing. Model M3 estimates that a one log-unit increase in downloading is associated with a 2.24 block increase in a user's repertoire. This supports the idea that remixing is a pathway to learning, in the broader conceptual sense of remixing that is used by many theorists. However, even with this strong control for downloads, this model estimates that a one log-unit increase in the number of remixes shared on Scratch is associated with a 0.25 block marginal increase in repertoire. Although reflecting only a portion of a block, this effect is well estimated ($\sigma=0.02$) and statistically significant even at the conservative $\alpha=0.001$ level.  Goodness of fit statistics shown in Table \ref{tab:rep} (e.g., $R^2=0.80$) suggest that these models fit the data well, and they explain a large majority of variation in Scratch users' cumulative repertoire of blocks.

The marginal effect sizes in these repertoire models are small, but can still be substantively meaningful. For example, M2 predicts that, holding all control variables at median values for similar users, a user who has shared 100 de novo projects -- but who had never remixed -- would be predicted to have a block repertoire of 81 blocks. If the user shared one remix for every three de novo projects, she would have a predicted repertoire of 86 blocks instead. When we control for downloads in M3, the estimated difference for this prototypical user is approximately one full block. Although small, this still reflects more than 1\% of the user's repertoire, and almost as large a proportion of all Scratch blocks.

\begin{table*}
\begin{center}

\begin{tabular}{l c c c c c c }
\hline
                      & Loops & Parallelism & Events & Conditionals & Operators & Data \\
\hline
ln Remixes w/ Concept & $0.388^{*}$  & $0.235^{*}$  & $0.282^{*}$  & $0.295^{*}$  & $0.158^{*}$
& $0.204^{*}$  \\
                      & $(0.009)$    & $(0.009)$    & $(0.009)$    & $(0.008)$    & $(0.007)$
& $(0.007)$    \\
ln Remixes            & $-0.283^{*}$ & $-0.406^{*}$ & $-0.355^{*}$ & $-0.463^{*}$ & $-0.433^{*}$
& $-0.530^{*}$ \\
                      & $(0.012)$    & $(0.012)$    & $(0.012)$    & $(0.010)$    & $(0.009)$
& $(0.009)$    \\
ln Comments           & $-0.586^{*}$ & $-0.286^{*}$ & $-0.574^{*}$ & $-0.728^{*}$ & $-0.702^{*}$
& $-0.657^{*}$ \\
                      & $(0.006)$    & $(0.006)$    & $(0.006)$    & $(0.005)$    & $(0.005)$
& $(0.005)$    \\
User Age              & $-0.006^{*}$ & $-0.006^{*}$ & $-0.010^{*}$ & $-0.011^{*}$ & $-0.024^{*}$
& $-0.026^{*}$ \\
                      & $(0.000)$    & $(0.000)$    & $(0.000)$    & $(0.000)$    & $(0.000)$
& $(0.001)$    \\
ln Experience         & $-0.152^{*}$ & $-0.286^{*}$ & $-0.191^{*}$ & $-0.100^{*}$ & $-0.061^{*}$
& $-0.045^{*}$ \\
                      & $(0.004)$    & $(0.004)$    & $(0.004)$    & $(0.003)$    & $(0.003)$
& $(0.003)$    \\
ln Total Blocks       & $0.283^{*}$  & $0.431^{*}$  & $0.330^{*}$  & $0.287^{*}$  & $0.306^{*}$
& $0.323^{*}$  \\
                      & $(0.003)$    & $(0.004)$    & $(0.004)$    & $(0.004)$    & $(0.004)$
& $(0.004)$    \\
ln Downloads          & $0.109^{*}$  & $0.033^{*}$  & $0.109^{*}$  & $0.068^{*}$  & $0.148^{*}$
& $0.139^{*}$  \\
                      & $(0.006)$    & $(0.006)$    & $(0.005)$    & $(0.005)$    & $(0.005)$
& $(0.005)$    \\
\hline
AIC                   & 963742   & 993339   & 908564   & 1193522  & 1231333
& 1152895  \\
R$^2$                 & 0.142        & 0.143        & 0.150        & 0.140        & 0.081
& 0.073        \\
Max. R$^2$            & 0.996        & 0.995        & 0.995        & 0.966        & 0.888
& 0.829        \\
Num. events           & 43615        & 44876        & 41122        & 51111        & 50658
& 46931        \\
Num. obs.             & 179522       & 190933       & 179061       & 370524       & 584950
& 681062       \\
Missings              & 4654         & 5415         & 5064         & 9148         & 15179
& 18393        \\
\hline
\multicolumn{7}{l}{\scriptsize{$^*p<0.001$}}
\end{tabular}

\caption{Fitted Cox proportional hazard models that estimate the ``risk'' that a Scratch user will share a de novo project that uses a block associated with a computational concept (e.g., loops) for the first time.}
\label{tab:surv}
\end{center}
\end{table*}

Our results for hypothesis H2 are shown in the six models described in Table \ref{tab:surv}. Although the relative size of the effects vary between concepts, the pattern of results are consistent. The marginal effect of the number of remixes with the CT concept is positive for all concepts. Parameter estimates can be interpreted as a ``magnifier'' of the hazard function. For example, a user who has remixed one log-unit more projects with conditionals would be estimated as being at 1.34 ($e^{0.29}$) times the risk of using conditionals in one of their de novo projects compared to an otherwise identical user.

\begin{figure*}[t]
\includegraphics[width=\textwidth]{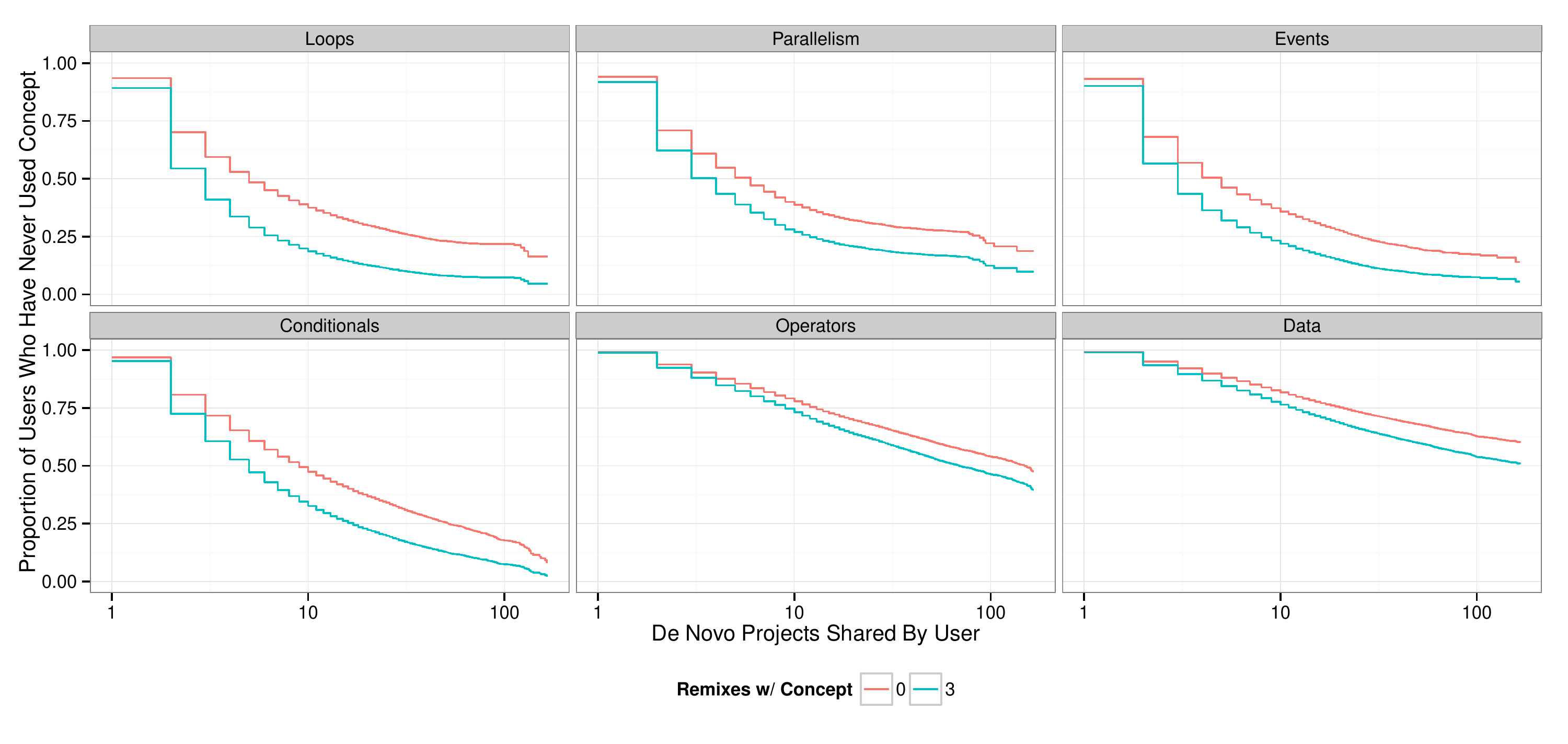}
\caption{Plots of model-derived estimates of the proportion of users who have used blocks associated with CT concepts for prototypical Scratch users. Estimates are shown for two prototypical users: (a) a user who has never shared a remix that uses the concept, and (b) a user has shared three such projects. All other variables are held at their sample median.}
\label{fig:protosurv}
\end{figure*}

To aid in interpretation, we include plots of model-derived estimates of the proportion of users who have used computational concepts for prototypical Scratch users in Figure \ref{fig:protosurv}. Estimates are shown for users from their first project up to 166 projects (the 99$\mathrm{th}$ percentile). Estimates are shown for two prototypical users: the first user has never shared a remix that uses the concept, while the second user has shared three remixes that do. All other variables are held at their overall sample median across all periods. The visualization shows that although the relative differences vary between concepts, Scratch users are systematically at higher ``risk'' of using blocks that demonstrate a given CT concept if they have remixed several projects that contain that concept.

With Cox models, one concern is that the assumption of proportional hazards may be violated if a variable has a stronger or weaker association at different points in ``time.'' Unfortunately, the test for proportionality is a function of the size of one's dataset, and our dataset is far too large for the test to be useful. To address whether our inference is affected, we examined Schoenfeld residual plots and found that average values are largely flat, although in some cases they appear to decrease toward zero as more projects are shared. Although we believe that a violation of the proportional hazards assumption is not affecting our fundamental inference, we caution readers to remain skeptical about the specific parameter estimates in the models.

Overall, our controls correspond to our predictions. \emph{Total Blocks} has a positive association with learning across our models and acts to attenuate the effects of other variables. Although previous work offered mixed predictions for the relationship between \emph{Comments} and learning, the results -- a positive association with repertoire but a negative association with concept use across the models in Table \ref{tab:surv} -- was unexpected and suggests a particularly interesting area for future research. Results for our controls for \emph{Age} and \emph{Experience} were also contrary to our predictions in the survival models. Of particular note are the estimates associated with \emph{Remixes} in Table \ref{tab:surv}. Once we control for remixes including a particular concept, the effect of the raw number of remixes is consistently negative. This is not surprising, since remixing projects that do not involve a concept seems unlikely to increase the likelihood that a particular concept will be used. We believe that the negative effect is caused by social forms of remixing like coloring-contests that are imperfectly measured by \emph{Total Blocks}.

Also worth noting is that \emph{Downloads} is consistently positive across all the models that include it. This reveals an interesting secondary result, as it suggests that simply looking at projects has a strong positive association with our measure of learning, and that Scratch may help support learning through less active forms of engagement. Alternatively, as discussed above, this positive association can also be seen as evidence for learning through a broader definition of remixing.

The results from both sets of models provide evidence that supports our original hypotheses. Our regression models of block repertoire in Table \ref{tab:rep} provide support for H1 and suggest that, taking the raw amount of code produced into account, users who engage in more remixing tend to have larger repertoires of programming blocks than users who engage in less. The survival models shown in Table \ref{tab:surv} and Figure \ref{fig:protosurv} point to support for H2, and suggest that remixing more projects with particular computational thinking concepts is associated with using those concepts in new projects. Among the six concepts, those with weaker associations with remixing in our data also showed lower adoption in general. This may indicate a potential lacuna in the learnability of concepts like data and operators, and warrants further investigation.

\section{Limitations}
The validity of our findings may be affected by a number of threats and limitations. One of these threats arises from the fact that our measures of learning may increase for reasons unrelated to, but correlated with, levels of remixing. To show that remixing \emph{causes} learning, we would need exogenous sources of variation in remixing which we do not have. This is a common limitation in learning research and in research on informal learning environments in particular. We have tried to address this concern by adding control variables that address the effects of likely confounds. Although we have included variables that attempt to control for users' popularity, experience, age, and technical sophistication, there may be other important controls that we have omitted.

A limitation of our hazard models is that a Scratch user may copy a stack of blocks that includes a block we are looking for without understanding the underlying concept behind the block. In robustness checks, we attempted to address this by using an alternative threshold for demonstration of a CT concept in each of our hazard models that ignores any demonstration of a CT concept if it occurs in a script that is identical to one used in a previous remix. Results are included in the supplemental material. With this alternate threshold, we discarded 2-15\% of de novo projects that were previously associated with a given CT concept. Because this criterion omits all pure cut-and-pasting by remixing users as evidence of learning, this acts as a conservative test that can only reduce the strength of our findings. In all six cases, the results are still positive, statistically significant, and only moderately reduce the parameter estimates for \emph{Remixes w/ Concept}.

Another threat stems from the fact that users join Scratch with different skills and initial repertoires. For example, a user who joins with a deeper initial knowledge base may grow more slowly but also be more likely to remix. In a related sense, Yang et.\ al.\ \cite{yang_uncovering_2015} take initial block repertoire into account when comparing user learning trajectories. To mitigate this threat, we added a control variable to our repertoire models that captures users' initial block repertoire. The addition of this control did not substantively impact our results.

A final threat is that the design of Scratch may make the CT concept of parallelism emerge unintentionally. For example, as Scratch projects often have multiple sprites, and since the standard way to start a Scratch project is to have an event-handler for ``when green flag clicked,'' a large number of multi-sprite projects may exhibit parallelism even though the user may not be able to use the concept more generally. To address this, we extended our algorithm to discard instances of parallelism between sprites, and only take into account parallel scripts that are within the same sprite. Results are included in the supplemental material. In line with our expectations, this specification strengthens the size of our parameter estimate as we discard possible unintentional parallelism.

Several of these threats hint at a broader, unavoidable limitation: no behavioral measure of learning can see inside users' minds to know what they have learned. In this sense, measuring learning in informal learning environments like Scratch is itself an open area of research which we believe this work contributes to. Although we believe that our measures of learning build and improve on previous work in this area (e.g., \cite{maloney_programming_2008, scaffidi_skill_2011, yang_uncovering_2015}), other metrics may lead to different results and conclusions.

Finally, like other work that studies activity within a single online community, questions of generalizability are both important and difficult to answer. Without further research, we cannot know if these findings generalize beyond the users in our sample. A more nuanced version of this limitation involves the fact that the large majority of Scratch users never share more than a very small number of projects. Scaffidi and Chambers' \cite{scaffidi_skill_2011} point that learning is unlikely to happen when most users engage only for short periods of time reflects both a methodological and substantive limitation that is critical to keep in mind when considering the impact and application of our findings.

\section{Discussion}
Our work finds support for the theory that users can learn through remixing. At the outset, we find that a Scratch user's repertoire of programming blocks is positively associated with the amount of remixing they engage in, controlling for the total amount of code a user is publishing. Diving deeper, we also find that users' demonstration of six key computational thinking concepts is positively associated with users' exposure to these concepts through remixing. We find our results are robust to a series of potential threats to validity. Although limited in a number of ways, these results provide support for proponents of remixing as a pathway to learning.

As discussed, the estimated effect sizes for our predictors are relatively small. Treating the relationship between \emph{Downloads} and learning as part of remixing's effect is suggestive of more substantive effects, but does not add up to a large effect itself. This is not entirely unexpected. Although Scratch was designed to promote learning through remixing, remixing itself is very unconstrained on the platform and has become dominated by genres like coloring contests which were unanticipated by Scratch's designers. Given the informal and messy nature of observational data from Scratch, we believe that our marginal effects still reveal important evidence of learning, and it is easy to imagine stronger effects in a more structured context.

For example, teachers could frame and construct remixing experiences so that learners are exposed to code-intense projects or to particular concepts. Brennan's work on combining formal learning environments with Scratch is one useful guide for future work in this direction \cite{brennan_best_2013}. Larger effects might also be achieved through increased structure within informal environments.  For example, research on other remixing platforms has suggested that large amounts of remixing activity can be encouraged and directed to particular source material by site administrators \cite{cheliotis_analysis_2009}. Even within Scratch, previous studies document efforts to promote ``high value'' forms of remixing through ``collaboration camps'' \cite{Roque:2012:TCD:2307096.2307130}.

Our work makes a series of contributions. Our most fundamental contribution is in testing the widespread and influential theory that remixing is positively associated with learning. Second, our paper builds on previous work to describe two novel methods for measuring computational thinking quantitatively. Because the Scratch dataset is becoming more widely available to researchers through a nascent public release, we hope our methods will provide a point of departure for future studies of Scratch. We also believe our approach can be adapted to other informal learning environments.

Of course, this work is only a first step. We hope to build on the current study by testing the relationship between our measures of learning and other predictors like levels and types of socialization. We hope that future work will critique and build on our theories, our measures, and our approach. As quantitative analysts of learning, we know that any attempt to measure learning is necessarily reductionist and potentially dangerous. We present our findings with the hope that others will build and improve upon our efforts.

Our findings have implications for designers interested in promoting learning, and our results support calls to incorporate remixing into the design of social computing systems. The differences between effect sizes in our hazard models suggest that remixing may be more effective at promoting engagement with some concepts, like loops, than others, like operators and data. This may be because the type of projects that are popular in the Scratch community tend to be those that leverage certain concepts over others.  Alternatively, these differences may point toward certain concepts being less understandable or learnable, at least within the context of Scratch. Both of these reflect opportunities for designers to foster a wider range of engagement. Our technique for measuring learning has immediate implications and usefulness for the designers of informal learning systems. For example, we hope to work with the Scratch team at MIT to use the measures in this study to evaluate the design of new Scratch features.

As researchers and designers attracted to the promise of remixing, we feel our results are a welcome validation of the idea that remixing can act as a pathway to learning. That said, promoting remixing requires prolonged engagement, and remains extremely difficult in informal learning environments. Moreover, promoting remixing among engaged users remains both difficult and fraught with unpalatable trade-offs \cite{hill_remixing_2013, cheliotis_antecedents_2014}. The fact that remixing may be associated with learning does not make it any easier to promote or any better along other dimensions. Although our results provide evidence that the promise seen in remixing may not be misplaced, enormous work remains to realize its potential.

\section{Acknowledgments}
We would like to thank the Lifelong Kindergarten group at the MIT Media Lab for creating Scratch, as well as the millions of Scratch users who create and participate on the Scratch website. We would also like to acknowledge Mitchel Resnick, Natalie Rusk, Samantha Shorey, Samuel Woolley, and our anonymous reviewers for their thoughtful feedback. Financial support for this work came from the National Science Foundation (grants DRL-1417663 and DRL-1417952).

\balance{}
\bibliographystyle{SIGCHI-Reference-Format}
\bibliography{refs}


\begin{thebibliography}{00}


\ifx \showCODEN    \undefined \def \showCODEN     #1{\unskip}     \fi
\ifx \showDOI      \undefined \def \showDOI       #1{{\tt DOI:}\penalty0{#1}\ }
  \fi
\ifx \showISBNx    \undefined \def \showISBNx     #1{\unskip}     \fi
\ifx \showISBNxiii \undefined \def \showISBNxiii  #1{\unskip}     \fi
\ifx \showISSN     \undefined \def \showISSN      #1{\unskip}     \fi
\ifx \showLCCN     \undefined \def \showLCCN      #1{\unskip}     \fi
\ifx \shownote     \undefined \def \shownote      #1{#1}          \fi
\ifx \showarticletitle \undefined \def \showarticletitle #1{#1}   \fi
\ifx \showURL      \undefined \def \showURL       #1{#1}          \fi

\bibitem{benkler_wealth_2006}
{Yochai Benkler}. 2006.
\newblock {\em The {Wealth} of {Networks}: {How} {Social} {Production}
  {Transforms} {Markets} and {Freedom}}.
\newblock Yale University Press.
\newblock
\showISBNx{0300110561}


\bibitem{brandt_two_2009}
{Joel Brandt}, {Philip~J. Guo}, {Joel Lewenstein}, {Mira Dontcheva}, {and}
  {Scott~R. Klemmer}. 2009.
\newblock \showarticletitle{Two studies of opportunistic programming:
  interleaving web foraging, learning, and writing code}. ACM Press, 1589.
\newblock
\showISBNx{9781605582467}
\showDOI{%
\url{http://dx.doi.org/10.1145/1518701.1518944}}


\bibitem{brennan_best_2013}
{Karen Brennan}. 2013.
\newblock {\em Best of both worlds: {Issues} of structure and agency in
  computational creation, in and out of school}.
\newblock Ph.D. Dissertation. Massachusetts Institute of Technology.
\newblock


\bibitem{brennan_new_2012}
{Karen Brennan} {and} {Mitchel Resnick}. 2012.
\newblock \showarticletitle{New frameworks for studying and assessing the
  development of computational thinking}. In {\em 2012 annual meeting of the
  {American} {Educational} {Research} {Association}, {Vancouver}, {Canada}}.
\newblock
\showURL{%
\url{http://scratched.gse.harvard.edu/ct/files/AERA2012.pdf}}


\bibitem{brennan_imagining_2013}
{Karen Brennan} {and} {Mitchel Resnick}. 2013.
\newblock \showarticletitle{Imagining, {Creating}, {Playing}, {Sharing},
  {Reflecting}: {How} {Online} {Community} {Supports} {Young} {People} as
  {Designers} of {Interactive} {Media}}.
\newblock In {\em Emerging {Technologies} for the {Classroom}}, {Chrystalla
  Mouza} {and} {Nancy Lavigne} (Eds.). Springer New York, 253--268.
\newblock
\showISBNx{978-1-4614-4695-8, 978-1-4614-4696-5}
\showURL{%
\url{http://link.springer.com/chapter/10.1007/978-1-4614-4696-5_17}}


\bibitem{brennan_more_2011}
{Karen Brennan}, {Amanda Valverde}, {Joe Prempeh}, {Ricarose Roque}, {Michelle
  Chung}, {Karen Brennan}, {Amanda Valverde}, {Joe Prempeh}, {Ricarose Roque},
  {and} {Michelle Chung}. 2011.
\newblock \showarticletitle{More than code: {The} significance of social
  interactions in young people's development as interactive media creators},
  Vol. 2011. 2147--2156.
\newblock
\showISBNx{1-880094-86-X}
\showURL{%
\url{http://www.editlib.org/p/38158/}}


\bibitem{bruckman_community_1998}
{Amy Bruckman}. 1998.
\newblock \showarticletitle{Community {Support} for {Constructionist}
  {Learning}}.
\newblock {\em Computer Supported Cooperative Work (CSCW)\/} {7}, 1 (March
  1998), 47--86.
\newblock
\showDOI{%
\url{http://dx.doi.org/10.1023/A:1008684120893}}


\bibitem{cheliotis_antecedents_2014}
{Giorgos Cheliotis}, {Nan Hu}, {Jude Yew}, {and} {Jianhui Huang}. 2014.
\newblock \showarticletitle{The {Antecedents} of {Remix}}. In {\em {ACM} 17th
  {Conference} on {Computer} {Supported} {Cooperative} {Work} \&\#38; {Social}
  {Computing}} {\em ({CSCW} '14)}. ACM, New York, NY, USA, 1011--1022.
\newblock
\showISBNx{978-1-4503-2540-0}
\showDOI{%
\url{http://dx.doi.org/10.1145/2531602.2531730}}


\bibitem{cheliotis_analysis_2009}
{Giorgos Cheliotis} {and} {Jude Yew}. 2009.
\newblock \showarticletitle{An analysis of the social structure of remix
  culture}. In {\em C\&{T} 2009}. 165--174.
\newblock
\showISBNx{978-1-60558-713-4}
\showDOI{%
\url{http://dx.doi.org/10.1145/1556460.1556485}}


\bibitem{dahotre_qualitative_2010}
{Aniket Dahotre}, {Yan Zhang}, {and} {Christopher Scaffidi}. 2010.
\newblock \showarticletitle{A {Qualitative} {Study} of {Animation}
  {Programming} in the {Wild}}. In {\em 2010 {ACM}-{IEEE} {International}
  {Symposium} on {Empirical} {Software} {Engineering} and {Measurement}} {\em
  ({ESEM} '10)}. ACM, New York, NY, USA, 29:1--29:10.
\newblock
\showISBNx{978-1-4503-0039-1}
\showDOI{%
\url{http://dx.doi.org/10.1145/1852786.1852825}}


\bibitem{fernando_online_2014}
{Champika Fernando}. 2014.
\newblock {\em Online learning webs : designing support structures for online
  communities}.
\newblock Thesis. Massachusetts Institute of Technology.
\newblock
\showURL{%
\url{http://dspace.mit.edu/handle/1721.1/95602}}


\bibitem{fiesler_remixers_2014}
{Casey Fiesler} {and} {Amy~S. Bruckman}. 2014.
\newblock \showarticletitle{Remixers' {Understandings} of {Fair} {Use}
  {Online}}. In {\em {ACM} 17th {Conference} on {Computer} {Supported}
  {Cooperative} {Work} \&\#38; {Social} {Computing}} {\em ({CSCW} '14)}. ACM,
  New York, NY, USA, 1023--1032.
\newblock
\showISBNx{978-1-4503-2540-0}
\showDOI{%
\url{http://dx.doi.org/10.1145/2531602.2531695}}


\bibitem{hartmann_hacking_2008}
{B. Hartmann}, {S. Doorley}, {and} {S.R. Klemmer}. 2008.
\newblock \showarticletitle{Hacking, {Mashing}, {Gluing}: {Understanding}
  {Opportunistic} {Design}}.
\newblock {\em IEEE Pervasive Computing\/} {7}, 3 (July 2008), 46--54.
\newblock
\showISSN{1536-1268}
\showDOI{%
\url{http://dx.doi.org/10.1109/MPRV.2008.54}}


\bibitem{hemphill_remix_2009}
{Scott~C. Hemphill} {and} {Jeannie Suk}. 2009.
\newblock \showarticletitle{Remix and {Cultural} {Production}}.
\newblock {\em Stanford Law Review\/} {61}, 1227 (2009).
\newblock


\bibitem{hill_cost_2013}
{Benjamin~Mako Hill} {and} {Andrés Monroy-Hernández}. 2013a.
\newblock \showarticletitle{The cost of collaboration for code and art:
  evidence from a remixing community}. In {\em {ACM} 2013 conference on
  {Computer} supported cooperative work} {\em ({CSCW} '13)}. ACM, New York, NY,
  USA, 1035--1046.
\newblock
\showISBNx{978-1-4503-1331-5}
\showDOI{%
\url{http://dx.doi.org/10.1145/2441776.2441893}}


\bibitem{hill_remixing_2013}
{Benjamin~Mako Hill} {and} {Andrés Monroy-Hernández}. 2013b.
\newblock \showarticletitle{The {Remixing} {Dilemma} {The} {Trade}-{Off}
  {Between} {Generativity} and {Originality}}.
\newblock {\em American Behavioral Scientist\/} {57}, 5 (May 2013), 643--663.
\newblock
\showISSN{0002-7642, 1552-3381}
\showDOI{%
\url{http://dx.doi.org/10.1177/0002764212469359}}


\bibitem{hill_responses_2010}
{Benjamin~Mako Hill}, {Andrés Monroy-Hernández}, {and} {Kristina Olson}.
  2010.
\newblock \showarticletitle{Responses to remixing on a social media sharing
  website}. In {\em Proc. {ICWSM} 2010}. AAAI, Washington, D.C., 74--81.
\newblock


\bibitem{ito_hanging_2009}
{Mizuko Ito} (Ed.). 2009.
\newblock {\em Hanging {Out}, {Messing} {Around}, and {Geeking} {Out}: {Kids}
  {Living} and {Learning} with {New} {Media}}.
\newblock The MIT Press.
\newblock


\bibitem{jenkins_convergence_2008}
{Henry Jenkins}. 2008.
\newblock {\em Convergence {Culture}: {Where} {Old} and {New} {Media}
  {Collide}\/} (revised ed.).
\newblock NYU Press, New York.
\newblock
\showISBNx{0814742955}


\bibitem{jenkins_confronting_2006}
{Henry Jenkins}, {Katie Clinton}, {Ravi Purushotma}, {Alice Robinson}, {and}
  {Margaret Weigel}. 2006.
\newblock {\em Confronting the {Challenges} of {Participatory} {Culture}:
  {Media} {Education} for the 21st {Century}}.
\newblock {T}echnical {R}eport. MacArthur Foundation, Chicago, Illinois, USA.
\newblock
\showURL{%
\url{http://digitallearning.macfound.org/site/apps/nlnet/content2.aspx?c=enJLKQNlFiG&b=2108773&content_id=}}


\bibitem{keen_cult_2007}
{Andrew Keen}. 2007.
\newblock {\em The {Cult} of the {Amateur}: {How} {Today}'s {Internet} is
  {Killing} {Our} {Culture}\/} (3rd printing ed.).
\newblock Crown Business.
\newblock
\showISBNx{0385520808}


\bibitem{kim_appropriate_2015}
{Sangmi Kim}, {Seong-Gyu Kim}, {Yoonsin Jeon}, {Soojin Jun}, {and} {Jinwoo
  Kim}. 2015.
\newblock \showarticletitle{Appropriate or {Remix}? {The} {Effects} of {Social}
  {Recognition} and {Psychological} {Ownership} on {Intention} to {Share} in
  {Online} {Communities}}.
\newblock {\em Human–Computer Interaction\/} {0}, ja (Feb. 2015).
\newblock
\showISSN{0737-0024}
\showDOI{%
\url{http://dx.doi.org/10.1080/07370024.2015.1022425}}


\bibitem{lanier_you_2010}
{Jaron Lanier}. 2010.
\newblock {\em You {Are} {Not} a {Gadget}: {A} {Manifesto}\/} (1 ed.).
\newblock Knopf.
\newblock
\showISBNx{0307269647}


\bibitem{lave_situated_1991}
{Jean Lave} {and} {Etienne Wenger}. 1991.
\newblock {\em Situated learning: {Legitimate} peripheral participation}.
\newblock Cambridge University Press.
\newblock
\showISBNx{0521413087}


\bibitem{lessig_remix:_2008}
{Lawrence Lessig}. 2008.
\newblock {\em Remix: {Making} {Art} and {Commerce} {Thrive} in the {Hybrid}
  {Economy}}.
\newblock Penguin Press HC.
\newblock


\bibitem{luther_edits_2010}
{Kurt Luther}, {Nicholas Diakopoulos}, {and} {Amy Bruckman}. 2010.
\newblock \showarticletitle{Edits \& credits: exploring integration and
  attribution in online creative collaboration}. In {\em {ACM} {CHI} 2010}.
  ACM, 2823--2832.
\newblock
\showISBNx{978-1-60558-930-5}
\showDOI{%
\url{http://dx.doi.org/10.1145/1753846.1753869}}


\bibitem{maloney_programming_2008}
{John~H Maloney}, {Kylie Peppler}, {Yasmin Kafai}, {Mitchel Resnick}, {and}
  {Natalie Rusk}. 2008.
\newblock \showarticletitle{Programming by choice: urban youth learning
  programming with {Scratch}}.
\newblock {\em ACM SIGCSE Bulletin\/} {40}, 1 (2008), 367--371.
\newblock


\bibitem{manovich_remix_2005}
{Lev Manovich}. 2005.
\newblock Remix and remixability.
\newblock   (2005).
\newblock
\showURL{%
\url{http://www.manovich.net/DOCS/Remix_modular.doc}}


\bibitem{monroy-hernandez_scratchr:_2007}
{Andrés Monroy-Hernández}. 2007.
\newblock \showarticletitle{{ScratchR}: sharing user-generated programmable
  media}. In {\em {ACM} 6th international conference on {Interaction} design
  and children} {\em ({IDC} '07)}. ACM, New York, NY, USA, 167--168.
\newblock
\showISBNx{978-1-59593-747-6}
\showDOI{%
\url{http://dx.doi.org/10.1145/1297277.1297315}}


\bibitem{monroy-hernandez_computers_2011}
{Andrés Monroy-Hernández}, {Benjamin~Mako Hill}, {Jazmin Gonzalez-Rivero},
  {and} {danah boyd}. 2011.
\newblock \showarticletitle{Computers can't give credit: how automatic
  attribution falls short in an online remixing community}. In {\em {ACM} 2011
  annual conference on {Human} factors in computing systems}. ACM, Vancouver,
  BC, Canada, 3421--3430.
\newblock


\bibitem{nickerson_appropriation_2011}
{Jeffrey~V. Nickerson} {and} {Andrés Monroy-Hernández}. 2011.
\newblock \showarticletitle{Appropriation and {Creativity}: {User}-{Initiated}
  {Contests} in {Scratch}}. In {\em 2011 44th {Hawaii} {International}
  {Conference} on {System} {Sciences} ({HICSS})}. 1--10.
\newblock
\showDOI{%
\url{http://dx.doi.org/10.1109/HICSS.2011.75}}


\bibitem{oehlberg_patterns_2015}
{Lora Oehlberg}, {Wesley Willett}, {and} {Wendy~E. Mackay}. 2015.
\newblock \showarticletitle{Patterns of {Physical} {Design} {Remixing} in
  {Online} {Maker} {Communities}}. In {\em {ACM} 33rd {Annual} {Conference} on
  {Human} {Factors} in {Computing} {Systems}} {\em ({CHI} '15)}. ACM, New York,
  NY, USA, 639--648.
\newblock
\showISBNx{978-1-4503-3145-6}
\showDOI{%
\url{http://dx.doi.org/10.1145/2702123.2702175}}


\bibitem{papert_poetic_1976}
{Seymour Papert}. 1976.
\newblock {\em Some {Poetic} and {Social} {Criteria} for {Education} {Design}}.
\newblock {T}echnical {R}eport 373. Massachusetts Institute of Technology,
  Cambridge, Massachusetts.
\newblock
\showURL{%
\url{http://dspace.mit.edu/handle/1721.1/6250}}


\bibitem{papert_mindstorms:_1980}
{Seymour Papert}. 1980.
\newblock {\em Mindstorms: {Children}, computers, and powerful ideas}.
\newblock Basic Books, Inc.
\newblock


\bibitem{peppler_uncovering_2011}
{Kylie~A Peppler} {and} {Mark Warschauer}. 2011.
\newblock \showarticletitle{Uncovering literacies, disrupting stereotypes:
  {Examining} the (dis) abilities of a child learning to computer program and
  read}.
\newblock  (2011).
\newblock


\bibitem{resnick_pianos_1996}
{Mitchel Resnick}, {Amy Bruckman}, {and} {Fred Martin}. 1996.
\newblock \showarticletitle{Pianos not stereos: {Creating} computational
  construction kits}.
\newblock {\em Interactions\/} {3}, 5 (1996), 40--50.
\newblock


\bibitem{resnick_scratch:_2009}
{Mitchel Resnick}, {John Maloney}, {Andrés Monroy-Hernández}, {Natalie Rusk},
  {Evelyn Eastmond}, {Karen Brennan}, {Amon Millner}, {Eric Rosenbaum}, {Jay
  Silver}, {Brian Silverman}, {and} {Yasmin Kafai}. 2009.
\newblock \showarticletitle{Scratch: programming for all}.
\newblock {\em Commun. ACM\/} {52}, 11 (2009), 60--67.
\newblock
\showDOI{%
\url{http://dx.doi.org/10.1145/1592761.1592779}}


\bibitem{Roque:2012:TCD:2307096.2307130}
{Ricarose Roque}, {Yasmin Kafai}, {and} {Deborah Fields}. 2012.
\newblock \showarticletitle{From Tools to Communities: Designs to Support
  Online Creative Collaboration in Scratch}. In {\em Proceedings of the 11th
  International Conference on Interaction Design and Children} {\em (IDC '12)}.
  ACM, New York, NY, USA, 220--223.
\newblock
\showISBNx{978-1-4503-1007-9}
\showDOI{%
\url{http://dx.doi.org/10.1145/2307096.2307130}}


\bibitem{scaffidi_skill_2011}
{Christopher Scaffidi} {and} {Christopher Chambers}. 2011.
\newblock \showarticletitle{Skill {Progression} {Demonstrated} by {Users} in
  the {Scratch} {Animation} {Environment}}.
\newblock {\em International Journal of Human-Computer Interaction\/} {28}, 6
  (June 2011), 383--398.
\newblock
\showISSN{1044-7318}
\showDOI{%
\url{http://dx.doi.org/10.1080/10447318.2011.595621}}


\bibitem{seneviratne_remix_2010}
{Oshani Seneviratne} {and} {Andrés Monroy-Hernández}. 2010.
\newblock \showarticletitle{Remix {Culture} on the {Web}: {A} {Survey} of
  {Content} {Reuse} on {Different} {User}-{Generated} {Content} {Websites}}. In
  {\em Web {Science}}. Raleigh, NC.
\newblock
\showURL{%
\url{http://journal.webscience.org/392/2/websci10_submission_109.pdf}}


\bibitem{singer_applied_2003}
{Judith~D. Singer} {and} {John~B. Willett}. 2003.
\newblock {\em Applied {Longitudinal} {Data} {Analysis}: {Modeling} {Change}
  and {Event} {Occurrence}\/} (1 ed.).
\newblock Oxford University Press, USA.
\newblock
\showISBNx{0195152964}


\bibitem{wing_computational_2006}
{Jeannette~M. Wing}. 2006.
\newblock \showarticletitle{Computational {Thinking}}.
\newblock {\it Commun. ACM} {49}, 3 (March 2006), 33--35.
\newblock
\showISSN{0001-0782}
\showDOI{%
\url{http://dx.doi.org/10.1145/1118178.1118215}}


\bibitem{wing_computational_2008}
{Jeannette~M. Wing}. 2008.
\newblock \showarticletitle{Computational thinking and thinking about
  computing}.
\newblock {\em Philosophical Transactions of the Royal Society of London A:
  Mathematical, Physical and Engineering Sciences\/} {366}, 1881 (Oct. 2008),
  3717--3725.
\newblock
\showISSN{1364-503X, 1471-2962}
\showDOI{%
\url{http://dx.doi.org/10.1098/rsta.2008.0118}}


\bibitem{wing_computational_2010}
{Jeannette~M Wing}. 2010.
\newblock Computational {Thinking} — {What} and {Why}?  (2010).
\newblock
\showURL{%
\url{http://www.cs.cmu.edu/~CompThink/resources/TheLinkWing.pdf}}


\bibitem{yang_uncovering_2015}
{Seungwon Yang}, {Carlotta Domeniconi}, {Matt Revelle}, {Mack Sweeney}, {Ben~U.
  Gelman}, {Chris Beckley}, {and} {Aditya Johri}. 2015.
\newblock \showarticletitle{Uncovering {Trajectories} of {Informal} {Learning}
  in {Large} {Online} {Communities} {Of} {Creators}}. In {\em {ACM} {Second}
  (2015) {Conference} on {Learning}@ {Scale}}. ACM, 131--140.
\newblock
\showISBNx{1450334113}


\bibitem{yew_social_2009}
{Jude Yew}. 2009.
\newblock \showarticletitle{Social performances: understanding the motivations
  for online participatory behavior}. In {\em {ACM} 2009 international
  conference on {Supporting} group work}. ACM, Sanibel Island, Florida, USA,
  397--398.
\newblock
\showISBNx{978-1-60558-500-0}
\showDOI{%
\url{http://dx.doi.org/10.1145/1531674.1531742}}


\bibitem{yu_cooks_2011}
{Lixiu Yu} {and} {Jeffrey~V. Nickerson}. 2011.
\newblock \showarticletitle{Cooks or cobblers?: crowd creativity through
  combination}. In {\em {ACM} 2011 annual conference on {Human} factors in
  computing systems} {\em ({CHI} '11)}. ACM, New York, NY, USA, 1393--1402.
\newblock
\showISBNx{978-1-4503-0228-9}
\showDOI{%
\url{http://dx.doi.org/10.1145/1978942.1979147}}


\end{thebibliography}

\end{document}